\documentclass[12pt]{article}

\usepackage{epsf,a4wide}

\newcommand{\sect}{Sect.}        
\newcommand{\fig}{Fig.}          

\newcommand{\eqpt}{\hspace{6pt}.\hspace{6pt}}  
\newcommand{\eqcm}{\hspace{6pt},\hspace{6pt}}
\newcommand{\GeV}{\mbox{\ GeV}}

\newcommand{\gp}{\gamma^\ast p}
\newcommand{\eps}{\varepsilon}      
\newcommand{\pom}{{I \! \! P}}      
\newcommand{\xp}{\xi}
\newcommand{\pdot}{\! \cdot \!}     

\newcommand{\lt}{\mathbf{l}_t}      
\newcommand{\dt}{\mathbf{\Delta}_t}

\newcommand{\PDL}{\mathcal{P}_{DL}^{\phantom{\ast \!\!}}}
\newcommand{\PLN}{\mathcal{P}_{LN}^{\phantom{\ast \!\!}}}

\newcommand{\fdag}[1]{\mbox{$\not\!#1$}}       
\newcommand{\deriv}
          {\mbox{$\stackrel{\scriptscriptstyle\leftrightarrow}{D^{}}$}}

\begin{document}

\begin{flushright}
  DAPNIA/SPhN--98--16 \\[2\baselineskip]
\end{flushright}

\begin{center}
{\Large THE DONNACHIE-LANDSHOFF POMERON \\[0.5\baselineskip]
AND GAUGE INVARIANCE} \\[2\baselineskip]
M. Diehl \\[0.1\baselineskip]
\textit{DAPNIA/SPhN, CEA/Saclay, 91191 Gif sur Yvette CEDEX, France}
\\[2\baselineskip]
\textbf{\large Abstract} \\[\baselineskip]
\parbox{0.9\textwidth}{\small The pomeron model of Donnachie and
Landshoff does not conserve the electromagnetic current when applied
to diffractive reactions such as electroproduction of a
quark-antiquark pair or of a vector meson. We propose a treatment of
this problem which ensures a physical behaviour of cross sections in
the photoproduction limit and show that it leads to results rather
similar to those obtained from two-gluon exchange.}
\end{center}

\vspace{0.5\baselineskip}

\section{Introduction}

The description of the pomeron within the framework of QCD remains one
of the great tasks in strong interaction physics. Important progress
has been made in this field over the last years, especially in the
study of diffractive processes with a hard scale, for instance a large
photon virtuality $Q^2$ in $ep$ collisions. On the other hand,
phenomenological models are still of interest in this new dynamical
regime: calculations in the framework of multi-gluon exchange can be
of considerable complexity, and it is useful to have models that
reproduce their results in a simpler way, thus allowing the
investigation of more complicated reactions with a reasonable amount
of effort. Furthermore, such models provide an opportunity to
extrapolate from the hard to the soft diffractive regime, e.g.\ to
take the limit of small $Q^2$ in diffractive $ep$ scattering, and to
make contact with elastic and diffractive hadron scattering, domains
where the application of QCD is much more difficult since perturbation
theory is of little help. It goes by itself that one cannot expect
such models to reproduce \emph{all} results and aspects of more
sophisticated and more fundamental QCD approaches.

In this paper we are concerned with the pomeron model of Donnachie and
Landshoff (DL)~\cite{DL-model} and its confrontation with the QCD
motivated model of nonperturbative gluon exchange by Landshoff and
Nachtmann (LN)~\cite{LN,DL-rho}. It was found in~\cite{MD} that the DL
model has problems with electromagnetic gauge invariance, which appear
for instance when it is applied to the production of a quark-antiquark
pair in diffractive $ep$ scattering. In \sect~\ref{sec:qqbar} we
explain the origin of this problem and propose a possible solution
which makes the DL model reproduce several, although not all results
of the two-gluon calculation in the LN approach. We then investigate
exclusive vector meson production under the same aspects. In
\sect~\ref{sec:disc} we discuss our way to handle the gauge invariance
problem from the point of view of contact terms, and make some
comparison with other approaches in the literature. We summarise our
results in \sect~\ref{sec:sum}.

\section{Diffractive $q \bar{q}$-production}
\label{sec:qqbar}

\subsection{The problem in the DL model and a solution}
\label{sec:problem}

In the DL model the pomeron couples to quarks via their vector
current, a choice originally motivated by the analysis of elastic and
diffractive hadron-hadron scattering~\cite{PVL-et-al}. In this sense
the pomeron is said to behave like a photon, \emph{except} that it has
charge conjugation parity $C = + 1$ unlike the photon with $C = -
1$. To implement this difference of quantum numbers the model
prescribes to subtract Feynman diagrams which are related by reversing
the charge flow of quark lines, instead of adding them as one would do
if the pomeron were replaced by a photon.

Calculating the forward $\gp$ scattering amplitude at high energy in
this model DL found that in order to obtain Bjorken scaling of the
proton structure function at small $x$ the quark-pomeron vertex cannot
be pointlike but must be softened at large quark
virtualities~\cite{DL-model}. They made the ansatz
\begin{equation}
\beta_0 \gamma^\mu \, f(k_1^2 - m_q^2) \, f(k_2^2 - m_q^2)
  \label{DL-coupl}
\end{equation}
with
\begin{equation}
f(k^2) = \mu_0^2 / (\mu_0^2 - k^2)
  \label{DL-formfactor}
\end{equation}
for the pomeron coupling to quarks of mass $m_q$ and momenta $k_1$ and
$k_2$, with the constants $\beta_0 \approx 2 \GeV^{-1}$ and $\mu_0
\approx 1 \GeV$ determined from phenomenology.

\begin{figure}
  \begin{center}
    \leavevmode
    \epsfsize=0.7\textwidth  \epsfbox{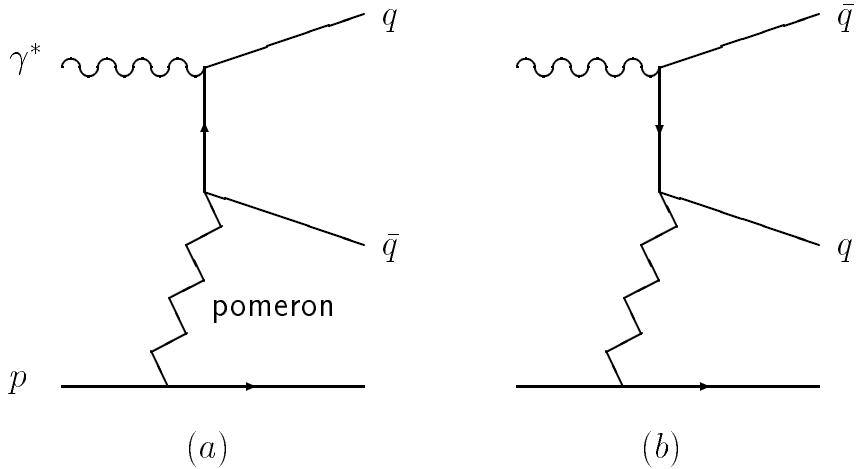}
  \end{center}
  \caption{\label{fig:diagrams-DL}Diffractive $q \bar{q}$-production
    in the DL model. The relative sign of the diagrams is the opposite
    of what it would be if the pomeron were replaced by a photon,
    because the pomeron and photon have opposite charge conjugation
    parity.}
\end{figure}

We now investigate diffractive $q \bar{q}$-production,
\begin{equation}
\gamma^\ast(q) + p(p) \to q(P_q) + \bar{q}(P_{\bar{q}}) + p(p')  \eqcm
  \label{reaction}
\end{equation}
with four-momenta given in parentheses. We will further use the
variables $\Delta = p - p'$, $t = \Delta^2$, $W^2 = (p + q)^2$, $Q^2 =
- q^2$, $M^2 = (P_q + P_{\bar{q}})^2$, $\beta = Q^2 / (2 \Delta \cdot
q)$ and $\xp = (\Delta \cdot q) / (p \cdot q)$. In the DL model this
reaction is described by the two diagrams of
\fig~\ref{fig:diagrams-DL}, taken with opposite relative sign. One
finds~\cite{MD} that the electromagnetic current is not conserved
here: the $\gp$ cross section $\sigma_U$ for the unphysical photon
polarisation $\eps_3^\mu = q_{\phantom{\!\!\!1}}^\mu /Q$ does not
vanish but instead has a $1 /Q^2$ singularity in the photoproduction
limit. In a technical sense the reason is that the two diagrams in
\fig~\ref{fig:diagrams-DL} have the ``wrong'' relative sign which was
assigned in order to implement the correct charge conjugation parity
of the pomeron---for photon instead of pomeron exchange gauge
invariance is of course guaranteed. If one uses the Feynman gauge for
the photon field and defines longitudinal and transverse photon
polarisations with respect to the $\gp$ axis in the c.m.\ of the
collision then the cross section $\sigma_L$ for longitudinal photons
has the same unphysical $1 /Q^2$ behaviour at small $Q^2$, only the
transverse cross section $\sigma_T$ is not affected by this problem.

It is instructive to look at the dependence of the cross sections on
the transverse momentum $p_t$ of the produced quark in the $\gp$ c.m.\
and on the invariant mass $M$ of the $q \bar{q}$-pair. Taking quark
masses equal to zero the unphysical and longitudinal cross sections
have a factor $p_t^2 /M^2$ relative to the transverse one, and as a
consequence one finds that at large $Q^2$ the $p_t$-integrated cross
sections $d \sigma_U /(d M^2)$ and $d \sigma_L /(d M^2)$ are
suppressed by $\mu_0^2 / Q^2 \cdot \log(Q^2 / \mu_0^2)$ compared with
$d \sigma_T /(d M^2)$. In this sense one can say that the gauge
violating terms are of higher twist and that the model can be used for
leading twist quantities. In jet production, where $p_t^2 /M^2$ is not
small, the unphysical cross section is however not negligible compared
with $d \sigma_T /(d p_t^2 \, d M^2)$, even at large $Q^2$, and the
model is of no use as it stands.

Let us have a closer look at the origin of the bad $Q^2$-behaviour of
$d \sigma_L /(d p_t^2 \, d M^2)$. From the expression of the
longitudinal polarisation vector in Feynman gauge
\begin{equation}
\eps_0  =  \frac{1}{\sqrt{1 + p^2 \, Q^2 / (p \pdot q)^2}} 
  \left( \frac{1}{Q} \, q + \frac{Q}{p \pdot q} \, p \right)
  \label{polaris-long}
\end{equation}
it is clear that the part of $\eps_0$ proportional to $p$ gives a
factor $Q$ in the amplitude and therefore a factor $Q^2$ in the $\gp$
cross section as required by gauge invariance. The unphysical
behaviour comes from the part proportional to $q$, which would give no
contribution if the electromagnetic current were conserved in the
model.

Without current conservation the results of a calculation will clearly
be gauge dependent. Therefore we need to \emph{fix} a photon gauge in
order to make the model well defined. We feel that this is legitimate
in the context of a phenomenological model, a choice of gauge then has
to be justified by its results. We choose to work in a noncovariant
gauge with gauge fixing vector $p$, where the photon field satisfies
$A \cdot p = 0$. In this gauge a polarisation vector $\eps$ in Feynman
gauge becomes
\begin{equation}
\eps \to \tilde{\eps} = \eps - \frac{p \pdot \eps}{p \pdot q} \, q
  \eqpt
  \label{polaris-modif}
\end{equation}
It is instructive to compare the tensor structure of the photon
propagator,
\begin{equation}
g^{\mu \nu} = - {\rm sgn}(q^2)\, \eps_0^\mu \eps_0^{\nu
\phantom{\!\!\!\mu}} - \eps_1^\mu \eps_1^{\nu \phantom{\!\!\!\mu}} -
\eps_2^\mu \eps_2^{\nu \phantom{\!\!\!\mu}} + {\rm sgn}(q^2)\,
\eps_3^\mu \eps_3^{\nu \phantom{\!\!\!\mu}} 
  \label{cov-gauge}
\end{equation}
in Feynman gauge and
\begin{equation}
g^{\mu \nu} - \frac{1}{p \pdot q} \left( p^\mu q^\nu + q^\mu p^\nu
\right) + \frac{p^2}{(p \pdot q)^2} \, q^\mu q^\nu = - {\rm sgn}(q^2)\,
  \tilde{\eps}_0^\mu \tilde{\eps}_0^{\nu \phantom{\!\!\!\mu}} -
  \tilde{\eps}_1^\mu \tilde{\eps}_1^{\nu \phantom{\!\!\!\mu}} -
  \tilde{\eps}_2^\mu \tilde{\eps}_2^{\nu \phantom{\!\!\!\mu}}
  \label{noncov-gauge}
\end{equation}
in our noncovariant gauge, where $\eps_1$ and $\eps_2$ are two
orthogonal vectors transverse to $p$ and $q$. We see that in the
amplitude for the electroproduction process $e p \to e p + q \bar{q}$
the term $\eps_3^\mu \eps_3^{\nu \phantom{\!\!\!\mu}}$ on the r.h.s.\
of (\ref{cov-gauge}) drops out when the index $\mu$ is contracted with
the electron current since this current is conserved, so that we are
left with the ill behaved contribution from the contraction of $q^\nu
/ Q$ in $\eps_0^\nu$ with the current of the produced quarks. In the
gauge~(\ref{polaris-modif}) the transverse polarisations, which
dominate in the $p_t$-integrated cross sections at large $Q^2$, are
the same as before, $\tilde{\eps}_1 = \eps_1$ and $\tilde{\eps}_2 =
\eps_2$, whereas the longitudinal polarisation
\begin{equation}
\tilde{\eps}_0 = \frac{1}{\sqrt{1 + p^2 \, Q^2 / (p \pdot q)^2}} \cdot
  \frac{Q}{p \pdot q} \left( p - \frac{p^2}{p \pdot q}\, q  \right)
  \label{polaris-long-modif}
\end{equation}
behaves like $Q$ in the photoproduction limit and leads to a
reasonable behaviour of the amplitude. From the l.h.s.\ of
(\ref{noncov-gauge}) we see that when changing from Feynman to our
noncovariant gauge we have effectively added the contraction of $-
p^\mu q^\nu / (p \cdot q)$ with the leptonic and hadronic currents to
the $e p \to e p + q \bar{q}$ amplitude, the terms with $q^\mu$ in the
photon propagator giving zero when contracted with the electron
current.

Instead of~(\ref{polaris-modif}) one may consider other noncovariant
gauges $A \cdot n = 0$ with a suitably chosen vector $n$. Introducing
longitudinal and transverse photon polarisations with respect to $n$
and $q$ and repeating the arguments that lead from
(\ref{polaris-long}) to (\ref{polaris-long-modif}) one will find that
the $ep$ amplitude has a well behaved photoproduction limit. This
amplitude will in general be different for different choices of $n$
since one has effectively added $- n^\mu q^\nu / (n \cdot q)$ to the
tensor $g^{\mu \nu}$ in the photon propagator. Note that the $\gp$
amplitude for photons that are transverse with respect to $p$ and $q$
will in general also be modified by such a choice of gauge.

A candidate gauge fixing vector is $n = \Delta$, which for $t = 0$ is
equivalent to $n = p$ since then $\Delta = \xp p$, but leads to
different results at finite $t$. Taking $n = \Delta$ looks more
symmetric with respect to the photon dissociation and proton
scattering parts of our reaction; to leading order in $\xp^{-1}$ the
choice $n = p$ is however equivalent to $n = p + q$, which is also
symmetric. We shall come back to the question of $p$ versus $\Delta$
in \sect~\ref{sec:rho}.

\subsection{Comparison of the DL and LN models}
\label{sec:compare}

We shall now see that in the gauge (\ref{polaris-modif}) the
predictions of the DL model for our process are remarkably similar to
those in the LN model, although they are not identical. We restrict
ourselves to $t = 0$ but admit a finite quark mass $m_q$ in the
produced $q \bar{q}$-pair. With Hand's convention for the flux factor
the cross section of the process (\ref{reaction}) for a given photon
polarisation reads
\begin{equation}
\left. \frac{d \sigma}{d t \, d p_t^2 \, d M^2} \right|_{t = 0} = 
    \frac{3}{128 \pi^3} \, \frac{\xp^2}{(Q^2 + M^2)^2 \, M^2
    \sqrt{1 - 4 (p_t^2 + m_q^2)/ M^2}} \,
    {\sum_{\mathit{spins}}}^{\! \prime} \, 
    \left| \mathcal{M}(\tilde{\eps}) \right|^2  \eqcm
\end{equation}
where $\sum_{\mathit{spins}}^{\, \prime}$ stands for spin summation for
the final state particles and spin average for the initial proton. In
the DL model one has
\begin{equation}
\mathcal{M}(\tilde{\eps}) = \frac{16 \pi e e_q}{3} \, \frac{j \pdot q}
  {2 p \pdot q} \, \xp^{-\alpha_\pom(0)} \cdot
  K \, \bar{u}(P_q) \, \gamma \pdot \tilde{\eps} \, v(P_{\bar{q}})
  \label{matrix-DL}
\end{equation}
with
\begin{equation}
K = \frac{9 \beta_0^2}{4 \pi} \, \frac{f(\hat{t} - m_q^2) + f(\hat{u} -
    m_q^2)}{2}  \eqpt
  \label{factor-DL}
\end{equation}
Here $\hat{t} = (q - P_q)^{2}$ and $\hat{u} = (q - P_{\bar{q}})^{2}$
are the Mandelstam variables of the pomeron-photon subreaction, $e_q$
is the charge of the produced quark in units of the positron charge
$e$ and $\alpha_\pom(t)$ is the soft pomeron trajectory.
\begin{equation}
j^\mu = \bar{u}(p') \left[ F_1(t) \gamma^\mu - \frac{i}{2 m_p}
\sigma^{\mu \nu} \Delta_\nu F_2(t) \right] u(p)
  \label{current}
\end{equation}
is the isoscalar nucleon vector current with $F_1$ and $F_2$ being the
isoscalar Dirac and Pauli nucleon form factors, i.e.\ the sum of the
respective form factors of proton and neutron.

\begin{figure}
  \begin{center}
    \leavevmode
    \epsfsize=0.7\textwidth  \epsfbox{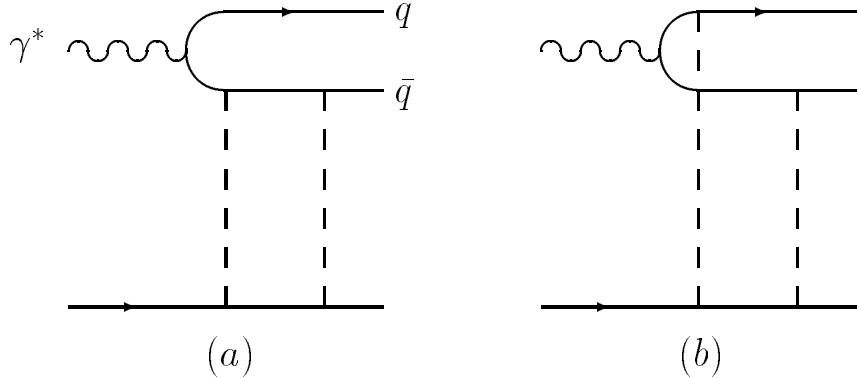}
  \end{center}
  \caption{\label{fig:diagrams-LN}Two of the four diagrams for $q
    \bar{q}$-production in the LN model, the other two are obtained by
    reversing the charge flow of the upper quark line. The lower line
    stands for a constituent quark in the proton and the pomeron is
    approximated by two nonperturbative gluons, represented by dashed
    lines.}
\end{figure}

In the LN model the reaction (\ref{reaction}) is described by the
diagrams of \fig~\ref{fig:diagrams-LN} plus those obtained by
reversing the charge flow of the upper quark line; for details of the
calculation cf.~\cite{MD}. Key quantities of the model are the moments
\begin{eqnarray}
\int_{0}^{\infty} d l^{2} [\alpha_{s}^{(0)} D(-l^{2})]^{2} &=&  
  \frac{9 \beta_{0}^{2}}{4\pi}  \nonumber \\
\int_{0}^{\infty} d l^{2} [\alpha_{s}^{(0)} D(-l^{2})]^{2} 
  \cdot l^{2} &=&  
  \frac{9 \beta_{0}^{2} \mu_{0}^{2}}{8\pi}    
\label{moments}
\end{eqnarray}
of the nonperturbative gluon propagator $D(l^2)$, where the strong
coupling in the nonperturbative region is taken as $\alpha_{s}^{(0)}
\approx 1$ following~\cite{alpha}. $\beta_0$ and $\mu_0$ have been
identified in~\cite{DL-rho} with the corresponding parameters in the DL
model, cf.~(\ref{DL-coupl}) and (\ref{DL-formfactor}). Note that
$\mu_0^2$ has here the significance of a typical gluon virtuality $l^2$
dominating in the integrals. For diffractive $q \bar{q}$-production we
have in our noncovariant gauge
\begin{eqnarray}
\mathcal{M}(\tilde{\eps}) &=& \frac{16 \pi e e_q}{3} \, 
  \frac{j.q}{2 p \pdot q} \, \xp^{-\alpha_\pom(0)} 
  \sqrt{ \frac{\alpha_s(\lambda^2)}{\alpha_s^{(0)}} }
  \cdot \left[ L_1 \, \bar{u}(P_q) \, \gamma \pdot
  \tilde{\eps} \, v(P_{\bar{q}}) \phantom{\frac{1}{2}}
  \right. \nonumber \\ 
&& {} - (L_1 - L_2) \, \frac{\tilde{\eps} \pdot q}{\Delta \pdot q} \,
  \bar{u}(P_q) \, \gamma \pdot \Delta \, v(P_{\bar{q}}) \nonumber \\ 
&& \left. {} + (L_1 - L_2) \, \frac{1}{\lambda^2} \, m_q \,
  \bar{u}(P_q) \, \gamma \pdot \tilde{\eps} \, \gamma
  \pdot \Delta \, v(P_{\bar{q}})  \right]  \eqcm
  \label{matrix-LN}
\end{eqnarray}
where
\begin{equation}
L_i = \int_{0}^{\infty} d l^2 \,
  [\alpha_{s}^{(0)} D(- l^2)]^{2} \, f_{i}(l^2, p_t^2, \lambda^2)
  \eqcm  \hspace{4em}  i = 1,2
  \label{loop-integrals}
\end{equation}
are loop integrals with weighting functions $f_1$, $f_2$ whose
expressions can be found in~\cite{MD}. The variable
\begin{equation}
\lambda^2 = \frac{p_t^2 + m_q^2}{1 - \beta}
  \label{Bartels-scale}
\end{equation}
can be seen as the relevant hard scale of the
process~\cite{Nikolaev-charm,BarLottWust}. The square root of
$\alpha_s(\lambda^2) /\alpha_s^{(0)}$ in (\ref{matrix-LN}) corresponds
to taking the quark-gluon coupling at a perturbative scale only for the
upper left vertex in the diagrams of \fig~\ref{fig:diagrams-LN}, where
one quark leg has a virtuality of order $\lambda^2$. This choice was
made in \cite{Cud} and \cite{MD,MD-angles,MD-charm}. One may argue that
the coupling should be taken as perturbative at both upper vertices,
then there is no square root in (\ref{matrix-LN}), (\ref{replace}). A
similar comment holds for vector meson production (\ref{LN-rho-matrix}).

At this point we remark that many features of the LN model are due to
the two-gluon exchange mechanism and in common with other two-gluon
approaches. The authors of \cite{BarLottWust} and \cite{Durham-charm}
use the gluon density in the proton to describe its coupling to the
two exchanged gluons; at $t = 0$ their expressions are related to
those in the LN model by the substitution\footnote{{}This substitution
is consistent with~\protect\cite{alpha}, where the gluon density was
estimated within the LN model.}
\begin{equation}
  \frac{\pi}{4} \alpha_s \cdot \frac{\partial}{\partial l^2}
  \left[ \xp\, g(\xp, l^2) \right]  \to
  \sqrt{ \frac{\alpha_s(\lambda^2)}{\alpha_s^{(0)}} } \cdot
  \xp^{1-\alpha_\pom(0)} \,
  [\alpha_{s}^{(0)} D(-l^{2})]^{2} \cdot l^{2} \eqcm
\label{replace}
\end{equation}
where $g(\xp, l^2)$ is the gluon density at a factorisation scale
$l^2$ and $\alpha_s$ on the l.h.s.\ taken at scale $\lambda^2$
in~\cite{BarLottWust} and $l^2$ in~\cite{Durham-charm}. The principal
difference in the predictions of the two approaches is thus the
$\xp$-dependence and overall normalisation, whereas the dependence on
$Q^2$, $M^2$, $p_t^2$ and the quark mass comes out very similar.

Coming back to the calculation in the LN model one finds that the
contribution of diagram $(a)$ in \fig~\ref{fig:diagrams-LN} plus the
one with reversed quark charge flow is obtained by replacing both
$L_1$ and $L_2$ with $9 \beta_0^2 / (4 \pi)$, so that only the first
term in the brackets of (\ref{matrix-LN}) survives. Except for the
root of $\alpha_s(\lambda^2) /\alpha_s^{(0)}$ this is just the DL
model expression (\ref{matrix-DL}), (\ref{factor-DL}) with the form
factor (\ref{DL-formfactor}) set equal to 1. In the DL model this form
factor ensures the decrease of the amplitude at large $p_t^2$, while
in the LN model the same effect is achieved by the contribution of
diagram $(b)$ and its analogue with opposite quark charge flow. In the
LN model one can of course also calculate with the polarisations
$\eps$ in Feynman gauge instead of $\tilde{\eps}$ since the sum of the
four diagrams is gauge invariant. If one does this~\cite{MD} then the
diagrams of type $(a)$ give contributions the amplitude for
longitudinal photons that do not vanish at $Q^2 \to 0$, which are
exactly cancelled by the diagrams of type $(b)$. In the DL model the
global factor $[f(\hat{t} - m_q^2) + f(\hat{u} - m_q^2)] /2$ cannot
achieve this and one is left with an unphysical behaviour at $Q^2 \to
0$ when working in Feynman gauge.

Comparing (\ref{matrix-DL}) with (\ref{matrix-LN}) we see that the
expression in the DL model can be obtained from the LN results by
replacing both loop integrals $L_1$ and $L_2$ by $K$ and by dropping
the ratio $\alpha_s(\lambda^2) /\alpha_s^{(0)}$ of strong couplings at
different scales in the LN expression. This ratio gives a different
normalisation as a function of the scale $\lambda^2$, whereas the
differences between the loop integrals (\ref{loop-integrals}) and the
combination of form factors (\ref{factor-DL}) have more subtle effects
as we shall now see.

At $t = 0$ the longitudinal polarisation vector reads
${\tilde{\eps}_0}^\mu = Q / (\Delta \cdot q) \cdot \Delta^\mu$ to
leading order in $\xp^{-1}$, and it is easy to see that the sum of the
first and the second term in the brackets of (\ref{matrix-LN}) gives
an amplitude proportional to $L_2$ for longitudinal photons. For
transverse photons the amplitude is proportional to $L_1$ in the case
$m_q = 0$ when only the first term contributes, and it involves both
loop integrals for finite quark mass due to the third term. These
differences are all absent in the DL result, where the effect of the
form factor is the same for all photon polarisations.

To illustrate the differences let us compare the values of $L_1$,
$L_2$ and $K$ in some kinematical limits, keeping in mind that there
is also an additional root of $\alpha_s(\lambda^2) / \alpha_s^{(0)}$
in (\ref{matrix-LN}). We always assume that $\mu_0^2 \ll Q^2 + M^2$ so
that from (\ref{DL-formfactor}), (\ref{factor-DL}) one has
\begin{equation}
K = \frac{9 \beta_0^2}{8 \pi} \, \frac{\mu_0^2}{\mu_0^2 + \lambda^2}
  \label{DL-simpler}
\end{equation}
up to terms suppressed by $\mu_0^2 / (Q^2 + M^2)$.

For $\lambda^2 \gg \mu_0^2$ a reasonable approximation of the loop
integrals $L_1$, $L_2$ is obtained by expanding the weighting
functions $f_1$, $f_2$ around $l^2 = 0$, and with (\ref{moments}) one
finds
\begin{equation}
    L_1 \approx \frac{9 \beta_0^2 \mu_0^2}{8 \pi \lambda^2} \cdot 2
    \left(1 - \frac{p_t^2}{\lambda^2} \right) \eqcm \hspace{1.5em} 
    L_2 \approx \frac{9 \beta_0^2 \mu_0^2}{8 \pi \lambda^2} \cdot 
    \left(1 - \frac{2 p_t^2}{\lambda^2} \right) \eqcm \hspace{1.5em} 
    K \approx \frac{9 \beta_0^2 \mu_0^2}{8 \pi \lambda^2}  \eqpt
    \label{lim-jets}
\end{equation}
In the case $p_t^2 \gg m_q^2$, which is relevant for jet production,
one has $\lambda^2 \approx p_t^2 / (1 - \beta)$ so that due to the
particular choice of form factor in (\ref{DL-formfactor}) the
$p_t^2$-dependence of the amplitude is the \emph{same} in the two
models for all photon polarisations. The $\beta$-dependence is however
different: in the LN model one finds a zero of the longitudinal
amplitude around $\beta = 1/2$ and a suppression of the transverse one
at small $\beta$, both phenomenologically important
effects~\cite{MD-angles,MD} which are not found in the DL model.

In the opposite limit, $p_t^2 \ll m_q^2$ at $\lambda^2 \gg \mu_0^2$,
which is relevant for heavy flavour production, one has $p_t^2 /
\lambda^2 \ll 1$ and thus $L_2 \approx K$. In the various $\gp$ cross
sections (eq.~(27) of \cite{MD-angles}) one finds that terms going with
$L_1$ are suppressed by powers of $p_t /m_q$ compared with those only
depending on $L_2$, so that in this limit the two models lead to the
same results, apart from the ratio of strong coupling constants.

When both $p_t$ and $m_q$ are small so that $\lambda^2 \ll \mu_0^2$ the
weighting functions $f_1$ and $f_2$ tend to 1 and we have
\begin{equation}
L_1, L_2 \approx \frac{9 \beta_0^2}{4 \pi} \eqcm \hspace{3em} 
K \approx \frac{9 \beta_0^2}{8 \pi}  \eqcm
  \label{lim-small-mass}
\end{equation}
so that in this limit the DL amplitude is just half of that in the LN
model, given that $\alpha_s(\lambda^2) / \alpha_s^{(0)}$ in the LN
expression should be replaced with 1 at small $\lambda^2$.

\subsection{The longitudinal diffractive structure function in the DL
model}
\label{sec:structure-fun}

As an application of our modification of the DL model we calculate the
integrated cross section for diffractive $q \bar{q}$-production, which
can be taken as an approximation for the inclusive diffractive cross
section---except in the region of small $\beta$ where diffractive states
with additional gluons are known to be important. Expressing our result
in terms of the diffractive structure functions $F^{D(4)}_T$ and
$F^{D(4)}_L$, conventionally defined by
\begin{eqnarray}
\frac{d \sigma_{ep}}{d x\, d Q^2\, d \xp\, d t} &=& \frac{4 \pi
  \alpha_{\it em}^2}{x Q^4} 
  \left[ \left( 1 - y + y^2/2 \right) F^{D(4)}_T + 
  (1 - y) \, F^{D(4)}_L \right]  \eqcm  \nonumber \\
F^{D(4)}_2 &=& F^{D(4)}_T + F^{D(4)}_L  \eqcm
  \label{diff-struct}
\end{eqnarray}
we obtain for $Q^2 + M^2 \gg \mu_0^2$
\begin{equation}
F^{D(4)}_{T,L}(\xp, \beta, Q^2, t) = f_\pom(\xp, t) \cdot
F^\pom_{T,L}(\beta, Q^2, t)
  \label{factorising}
\end{equation}
with
\begin{equation}
f_\pom(\xp, t) =  \frac{9 \beta_{0}^{2}}{4 \pi^{2}} \,
        \xp^{1 - 2 \alpha_{\pom}(t)} 
        \left[ F_1(t)^2 - \frac{t}{4 m_p^2} F_2(t)^2 \right]
  \label{DL-flux}
\end{equation}
and
\begin{eqnarray}
F_T^\pom(\beta, Q^2) &=& \frac{4}{3} \cdot \frac{3 \beta_0^2
  \mu_0^2}{8 \pi^2} \left\{ \beta (1 - \beta) + 
  O\left( \frac{\beta \mu_0^2}{Q^2} \right) \right\}
  \eqcm \nonumber \\
F_L^\pom(\beta, Q^2) &=& \frac{4}{3} \cdot \frac{3 \beta_0^2
  \mu_0^2}{8 \pi^2} \left\{ 4 \beta^3 \, \frac{\mu_0^2}{Q^2}
  \left[ \ln\left(\frac{Q^2}{\beta \mu_0^2}\right) - 1 + \frac{\beta
  \mu_0^2}{Q^2} \right]
  + O\left( \left(\frac{\beta \mu_0^2}{Q^2}\right)^3 \right) \right\}
  \eqcm
  \label{DL-long-trans}
\end{eqnarray}
where we have summed over massless quarks $u, \bar{u}, d, \bar{d}, s,
\bar{s}$. We have made the approximation $t = 0$ in $F_T^\pom$ and
$F_L^\pom$ but kept the strong $t$-dependence from the pomeron
trajectory and the elastic form factors in $f_\pom$. The terms of
higher order in (\ref{DL-long-trans}) give only small corrections: we
find that the effect of the terms of order $\beta \mu_0^2 / Q^2$ in
$F^\pom_T$ is below 3\% for $Q^2 / \beta \ge 5 \GeV^2$ and that the
approximation for $F^\pom_L$ is better than 4\% for $Q^2 / \beta \ge
10 \GeV^2$.

The result for $F_T^\pom$ is what has long been obtained by DL
\cite{DL-model}. It shows a scaling behaviour while the longitudinal
structure function is power suppressed in $1 / Q^2$, with $Q^2 \cdot
F^\pom_L$ having only a weak logarithmic dependence on $Q^2$. As
$\beta \to 1$ the transverse structure function vanishes like $1 -
\beta$ whereas the longitudinal one goes to a constant. All these
features are also found in the LN model~\cite{MD-charm}. In the DL
model the ratio $F^\pom_L / F^\pom_T$ turns out to be numerically
rather big even at moderately large $\beta$: for $Q^2$ between $10
\GeV^2$ and $30 \GeV^2$ it becomes equal to 1 at $\beta$ between
$0.65$ and $0.8$. Taking $Q^2 = 25 \GeV^2$ one finds $F^\pom_2$ rising
all the way up to $\beta = 1$, with the increase of $F^\pom_L$
overcompensating the decrease of $F^\pom_T$ for $\beta > 1/2$. This
looks difficult to reconcile with what is seen in the HERA data
\cite{H1-struct}.

In the LN model one also finds that $F^\pom_L$ becomes numerically
important at large $\beta$ where $F^\pom_T$ goes to
zero~\cite{MD-charm}, but there the effect is much less dramatic. This
is because of the extra factor of the running strong coupling at the
scale $\lambda^2 = p_t^2 / (1 - \beta)$. It leads to a suppression at
large $\beta$ which is stronger for $F^\pom_L$ than for $F^\pom_T$
since the former is less dominated by small $p_t^2$ than the
latter. It thus seems that the scale dependence of the strong coupling
which appears in the two-gluon calculation leads to a more realistic
prediction for the longitudinal structure function.

\section{Exclusive production of vector mesons}
\label{sec:rho}

The first process for which the DL and LN models have been compared
and where their close relation was observed~\cite{DL-rho} was elastic
electroproduction of a vector meson $V = \rho, \phi, J/\psi, \ldots$,
\begin{equation}
\gamma^\ast(q) + p(p) \to V(q') + p(p')  \eqpt
\end{equation}
It turns out that the DL model in its original form also violates
electromagnetic current conservation in this process, as was already
remarked in~\cite{Schaefer}. For the $\gp$ cross section one obtains
\begin{equation}
  \frac{d \sigma}{d t} = 
    \frac{1}{16 \pi\, W^4}\, {\sum_{\mathit{spins}}}^{\! \prime} \, 
    \left| \mathcal{M} \right|^2
\end{equation}
with
\begin{equation}
\mathcal{M} = \frac{48 M^2}{e}\, 
    \sqrt{\frac{3\pi \Gamma_{e^+ e^-}}{M}}\,
    \frac{j \pdot q}{2 p \pdot q} \,
    \xp^{- \alpha_\pom(t)}\, \PDL(\eps,\eps') \cdot 
    \frac{\beta_0^2 \mu_0^2}{Q^2 + M^2 - t + 2 \mu_0^2}  \eqcm
  \label{DL-rho-matrix}
\end{equation}
where $M$ denotes the meson mass, $\Gamma_{e^+ e^-}$ its decay width
into $e^+ e^-$, and $\eps$ and $\eps'$ the respective polarisation
vectors of photon and meson. The polarisation dependence is given by
\begin{equation}
\PDL(\eps,\eps') = 
   \frac{(p \pdot \eps)\, (\Delta \pdot \eps') -
   (\Delta \pdot \eps)\, (p \pdot \eps') +
   (\eps \pdot \eps')\, (p \pdot q)}{p \pdot q}  \eqcm
 \label{DL-rho-polar}
\end{equation}
where we have used $\eps' \cdot q' = 0$ but not made any assumption
about $\eps$ so that this expression holds in any photon gauge. As
announced $\PDL$ does not vanish for the unphysical polarisation $\eps =
\eps_3$. Working in the collision c.m.\ and defining polarisation
vectors with respect to the $p$\,--\,$q$ axis for the photon and with
respect to the $p'$\,--\,$q'$ axis for the meson we find in particular
that for the transition of a longitudinal photon to a longitudinal
vector meson one has
\begin{equation}
\PDL(L,L) = \frac{Q^2 - M^2 + t}{2 M Q}
  \label{DL-rho-long}
\end{equation}
in Feynman gauge, with an unphysical behaviour at small $Q$. In the LN
model one finds\footnote{{}Our expression for $\mathcal{M}$ differs from
those in \protect\cite{DL-rho} and \protect\cite{Cud,CudellRoyen} by
numerical factors. Making the replacement (\protect\ref{replace}) we
agree with the result of \cite{Durham-psi}.}
\begin{eqnarray}
\lefteqn{ \hspace{-1em} \mathcal{M} = \frac{48 M^2}{e}\, 
    \sqrt{\frac{3\pi \Gamma_{e^+ e^-}}{M}}\,
    \frac{j \pdot q}{2 p \pdot q} \,
    \xp^{- \alpha_\pom(t)} \,
    \sqrt{ \frac{\alpha_s(\lambda^2)}{\alpha_s^{(0)}} } \,\,
    \PLN(\eps,\eps') \, \cdot }
    \nonumber \\
 && \hspace{-1.5em} \frac{8}{9} \int d^2 l_t \,
    [ \alpha_{s}^{(0)} ]^2 
    D\left((\lt - \dt /2)^2\right) D\left((\lt + \dt /2)^2\right) 
    \frac{(\lt^2 - \dt^2 /4)}{Q^2 + M^2 - t + 4 (\lt^2 - \dt^2 /4)}
    \eqcm  \label{LN-rho-matrix}
\end{eqnarray}
where $\Delta_t$ is the transverse part of $\Delta$ with respect to
$p$ and $q$, and
\begin{equation}
\PLN(\eps,\eps') = 
   \frac{(p \pdot \eps)\, (\Delta \pdot \eps') -
   (\Delta \pdot \eps)\, (p \pdot \eps') +
   (\eps \pdot \eps')\, (p \pdot q) +
   \xp \, (p \pdot \eps)\, (p \pdot \eps')}{p \pdot q}  \eqpt
 \label{LN-rho-polar}
\end{equation}
Again this expression is valid in any photon gauge. If we require $Q^2
+ M^2 \gg \mu_0^2$ and remember that the typical gluon virtualities in
the integrals~(\ref{moments}) are of order $\mu_0^2$ then the terms
after $\PDL$ and $\PLN$ in (\ref{DL-rho-matrix}) and
(\ref{LN-rho-matrix}) are equal at $t = 0$; their difference will only
become important when $-t$ is of order $\mu_0^2$. Apart from this and
from the root of $\alpha_s(\lambda^2) / \alpha_s^{(0)}$, where in
analogy to (\ref{Bartels-scale}) one would now choose $\lambda^2 =
(Q^2 + M^2) /4$, the amplitudes in the two models then differ only by
their polarisation factors $\PDL$ and $\PLN$. The latter has an extra
term $\xp \, (p \cdot \eps)\, (p \cdot \eps')$ in the numerator, which
was overlooked in~\cite{DL-rho,Cud} and recently reported
in~\cite{CudellRoyen}. It precisely restores gauge invariance and
guarantees a reasonable small-$Q^2$ behaviour
\begin{equation}
  \PLN(L,L) = \frac{Q}{M}  
 \label{LN-rho-long}
\end{equation}
for the transition from a longitudinal photon to a longitudinal meson.
(\ref{DL-rho-long}) and (\ref{LN-rho-long}) do not even agree in the
limit $Q^2 \gg M^2$, where $\PLN(L,L)$ is twice as large as
$\PDL(L,L)$. For transverse photons on the other hand $\PDL$ gives the
same as $\PLN$. Note however that, unlike in the case of the
diffractive structure function discussed in
\sect~\ref{sec:structure-fun}, transverse photon polarisation in
vector meson production is suppressed at large $Q^2$; with
(\ref{LN-rho-polar}) one finds $\sigma_L / \sigma_T = Q^2 / M^2$ for
the ratio of longitudinal and transverse cross sections at all $Q^2$
and $t$.

If we work in the gauge $A \cdot p = 0$ then the extra term in $\PLN$
vanishes identically so that $\PDL$ agrees with $\PLN$ for all physical
photon polarisations. Defining the DL model in this gauge we thus find
that for $Q^2 + M^2 \gg \mu_0^2$ and $-t \ll \mu_0^2$ it gives the same
result as the LN approach up to the scale dependent ratio
$\alpha_s(\lambda^2) / \alpha_s^{(0)}$. This corresponds to what we
found for $q \bar{q}$-production in the limit $m_q^2 \gg p_t^2$ and
$\lambda^2 \gg \mu_0^2$ in \sect~\ref{sec:compare}; note that
following~\cite{DL-model} in the calculation of meson production we have
used a constituent mass of $M/2$ for the quarks and neglected their
transverse momentum in the meson so that kinematically the two processes
are equivalent.

Let us finally compare the two gauges $A \cdot n = 0$ with $n = p$ and
$n = \Delta$ at finite $t$. With $\Delta$ as gauge fixing vector one
does \emph{not} reproduce the results of the LN calculation since the
extra term in $\PLN$ does not always vanish. This happens when the
photon is transversely polarised in the scattering plane while the
meson is longitudinal. $\PLN$ is then negligible in the small-$\xp$
limit whereas $\PDL = - \sqrt{|\,t\,|}\, / M$ is not: the DL model
defined in this gauge thus violates $s$-channel helicity
conservation. Both from this point of view and in order to reproduce
as closely as possible the two-gluon result the choice $n = p$
therefore seems preferable to us.

\section{Discussion}
\label{sec:disc}

\subsection{Contact terms}
\label{sec:contact}

The point of view underlying our detailed comparison of the DL and the
LN models is that diffraction can be described in QCD by multi-gluon
exchange in the colour singlet channel, and that the LN model is a
simple implementation of this idea. The DL model, on the other hand,
looks more like an effective theory in which gluon degrees of freedom
are no longer explicitly present.

It can be shown that in the high-energy limit the coupling of two
$t$-channel gluons to a single quark line can be written in terms of
the quark vector current~\cite{LN}. Thus the diagram in
\fig~\ref{fig:diagrams-LN} $(a)$ is equivalent to the one in
\fig~\ref{fig:diagrams-DL} $(a)$ calculated without a form factor for
the quark-pomeron coupling, as we reported in the sequel of
(\ref{replace}). Technically this is seen as follows: for the upper
quark line in \fig~\ref{fig:diagrams-LN} $(a)$ one has $\fdag{p} \,
(\fdag{k} + m_q) \fdag{p} \approx 2 (p \cdot k) \fdag{p}$ to leading
order in energy, where $k$ denotes the momentum of the quark line
between the two quark-gluon vertices, while $p$ is the dominant part
of the gluon polarisation at each vertex and becomes the dominant
``pomeron polarisation'' in the DL model. One can also explicitly see
how this gives the correct $C = +1$ quantum number of the exchange:
reversing the charge flow of the quark line but keeping the flow of
its momentum unchanged one obtains $\fdag{p} \, (-\fdag{k} + m_q)
\fdag{p} \approx - 2 (p \cdot k) \fdag{p}$\,; the relative minus sign
is precisely what is put in by hand between the two diagrams in the DL
model.

The diagram in \fig~\ref{fig:diagrams-LN} $(b)$ has a different
topology than the diagrams of \fig~\ref{fig:diagrams-DL}, and its
contribution to the amplitude cannot be entirely rewritten in terms of
an effective quark-pomeron coupling with a form factor. It is
plausible to assume that a part of its contribution will have the
structure of a contact interaction between the quark line, the pomeron
and the photon; note that the generation of contact interactions is
well known in effective field theories. Such a contact interaction can
have a rather rich structure and we are not attempting here to
construct a corresponding extension of the DL model, in which gauge
invariance is restored. Let us however make the hypothesis that such
an extension can be found, where the sum of all diagrams for a process
conserves the electromagnetic current. Terms with an unphysical
$1/Q$-behaviour from the diagrams of \fig~\ref{fig:diagrams-DL} will
then be cancelled by terms from the diagram with a
quark-pomeron-photon contact interaction.

Now the procedure of gauge fixing we proposed in
\sect~\ref{sec:problem} can be seen in a different light. In a gauge
invariant theory one can of course work in \emph{any} photon gauge;
some terms in the amplitude will be ``shifted'' between different
diagrams when the gauge is changed. There will be gauges for which the
diagrams with and without contact term are separately well behaved in
$Q$; an example is just the gauge $A \cdot n = 0$ with a suitably
chosen vector $n$. Leaving out the contribution from the contact term
(because we do not know its form) we then make an error, but this
error is finite in the photoproduction limit. The error will be
different for different choices of $n$; and whether there is a
particularly good choice or even one where the contact term is
completely eliminated we can of course not determine without knowing
the specific form of the contact interaction. Nevertheless there are
certain criteria: our discussion at the end of \sect~\ref{sec:rho}
shows for instance that there are ``unfortunate'' choices for which
$s$-channel helicity in meson production is not conserved by the
diagrams without contact terms; in a model reproducing the two-gluon
exchange results such helicity violating terms will then only be
cancelled when contact interaction diagrams are taken into account.

\subsection{Comparison with other approaches}
\label{sec:other}

Several approaches to the problem of coupling the pomeron to quarks
have been made in the literature. In~\cite{Kramer} and later
in~\cite{Gehrman} the pomeron was treated exactly like a photon, with
a $\gamma^\mu$-coupling but without a form factor and without any
relative minus signs between diagrams. By construction this model does
not have any trouble with gauge invariance, but at the price of
actually describing $C = -1$ instead of $C = +1$ exchange. It can for
instance not describe the forward $\gp$ scattering amplitude, i.e.\
the inclusive proton structure function at small $x$, since the
coupling of three vector particles (two $\gamma^\ast$ and the pomeron)
to a quark loop vanishes due to Furry's theorem. For diffractive $q
\bar{q}$-production it gives a $p_t$-behaviour that is remarkably
different from the one in the DL calculation.  The DL model with the
form factor (\ref{DL-formfactor}), as well as the LN model, gives a
transverse $\gp$ cross section $d \sigma_T /(d p_t^2 \, d M^2)$ that
falls off approximately like $1 /p_t^4$ at large $p_t$ and is finite
as $p_t$ goes to zero. In the approach of~\cite{Kramer,Gehrman} the
decrease at large $p_t$ is much slower, and for massless quarks one
has a $1 /p_t^2$-singularity at zero $p_t$. This has important
consequences for the description of the leading twist part of $F^D_2$
in terms of diffractive parton densities, whose evolution equation
then has an \emph{inhomogeneous} term---just as in the case of parton
densities in the photon---which is intimately related with this
collinear singularity in the transverse $q \bar{q}$ cross
section. Such an inhomogeneous term will not be generated in the DL
model. We also remark that unlike the DL and LN approaches the pomeron
model of~\cite{Kramer,Gehrman} gives a leading twist contribution to
the longitudinal structure function $F^D_L$, going like $\beta ^2 (1 -
\beta)$, in strict analogy with the photon structure
function~\cite{Witten}.

The model of~\cite{Madrid} assumes a pointlike quark-pomeron coupling
through the scalar instead of the vector current of the quarks, i.e.\
it treats the pomeron as a scalar exchange. This is again gauge
invariant, and the scalar current has the correct quantum numbers to
model the pomeron, in particular it is $C = +1$. Taking the produced
quarks as massless one obtains a transverse $q \bar{q}$ cross section
going like $ 1 / p_t^2$ at large and at small $p_t$---the divergence
will again lead to an inhomogeneous evolution equation for diffractive
quark densities---and a longitudinal $q \bar{q}$ cross section that
vanishes at $t = 0$. We note that for massless quarks a scalar
quark-pomeron coupling flips the quark helicity, in contrast to models
with a vector coupling which conserves it. The helicity of massless
quarks is of course also conserved by multi-gluon exchange, at least
if one considers the $\gamma^\mu$-structure of the perturbative
quark-gluon vertex to be relevant in this context.

The starting point of~\cite{Schaefer} is a pomeron with a vector
coupling and a sign factor to ensure $C = +1$ exchange; a scalar term
accompanied with specific sign instructions is then added to the
$\gamma^\mu$-part of the coupling in order to restore current
conservation in vector meson production.  For a programme of
investigating the general structure of the pomeron-quark vertex we
refer to~\cite{Golosk}.

An effective theory for high energy quark-quark scattering, starting
from multi-gluon exchange, has been formulated in~\cite{Nacht}. The
current for $C = +1$ exchange has the structure $\bar{\psi} \gamma^\mu
\deriv_{\nu_1} \deriv_{\nu_2} \ldots \deriv_{\nu_{2n+1}} \psi$, where
the odd number of covariant derivatives gives the correct quantum
numbers of the pomeron. The Dirac matrix $\gamma^\mu$ leads to the
same quark spin structure of the coupling as in the DL model, while
the derivatives guarantee the correct signs between diagrams when the
quark charge flow is reversed, in a manner similar to the explicit
example of two-gluon exchange discussed in the previous subsection.

\section{Summary}
\label{sec:sum}

There are several possibilities to describe the pomeron coupling to
quarks in a phenomenological model. The Donnachie-Landshoff model
chooses a $\gamma^\mu$-coupling and implements the $C = +1$ parity of
the pomeron by introducing appropriate relative minus signs between
diagrams. In order to obtain phenomenologically reasonable results the
pomeron-quark coupling \emph{cannot} be pointlike in this model. We
point out that multi-gluon exchange in QCD motivates a form factor
behaviour of the quark-pomeron coupling\footnote{{}This should at
least hold for the part of the coupling that gives the leading twist
contribution to the diffractive structure function
$F_2^D$. In~\protect\cite{Bartels-twist} a possible pointlike
structure of the pomeron in perturbative QCD was discussed at the
level of \emph{nonleading} twist.} as well as helicity conservation
for massless quarks, which is ensured by the $\gamma^\mu$-vertex in
the DL model.

The price to be paid for introducing minus signs ``by hand'' in this
model is that the electromagnetic current is not conserved. We propose
to define the model by specifying a noncovariant gauge $A \cdot n = 0$
in which a good $Q^2$-behaviour is guaranteed for all physical photon
polarisations in the limit where the photon becomes real. The choice
of a gauge fixing vector $n$ is not unique and has to be motivated,
for instance by comparing predictions with more elaborate models such
as two-gluon exchange. Our preferred choice here is $n = p$.

A way to restore current conservation is the introduction of contact
terms between quark, photon and pomeron lines, which seems natural if
one thinks of the model as an ``effective theory'' of multi-gluon
exchange. To construct such terms is beyond the scope of this work,
but we note that in such a model one can make a suitable choice of
photon gauge to minimise the contribution of contact term graphs, so
that reasonable results may be obtained even when they are left
out. Our choice of gauge then assures in particular a good
$Q^2$-behaviour of the diagrams with and without contact terms
separately.

We have applied the DL model in the gauge $A \cdot p = 0$ to
diffractive $q \bar{q}$-production at $t = 0$ and compared in some
detail its results with those of the LN model of two-gluon
exchange. While their predictions differ in detail there is a strong
similarity between the two models, in particular for the
$p_t$-dependence, which leads to the prediction that $F^D_T$ is of
leading twist whereas $F^D_L$ is not. For vector meson production we
find almost identical results in the two models when $Q^2$ or the
meson mass is large and $t$ is small, provided one takes $p$ as gauge
fixing vector. Taking $\Delta$ instead would lead to a violation of
$s$-channel helicity conservation, at variance with two-gluon
exchange.

\section*{Acknowledgments}

I gratefully acknowledge discussions with A. Donnachie, J.-R. Cudell,
A. Hebecker, B. Kniehl, G. Kramer, J.-M. Laget, P. V. Landshoff,
O. Nachtmann, O. Teryaev and J. Vermaseren. Special thanks go to
J.-M. Laget and P. V. Landshoff for reading the manuscript. This work
has been partially funded through the European TMR Contract
No.~FMRX--CT96--0008: Hadronic Physics with High Energy
Electromagnetic Probes.

\end{document}